\documentclass[12pt]{article}
\usepackage{amssymb}
\usepackage{amsmath}
\usepackage{IEEEtrantools}
\usepackage{pdflscape}
\usepackage{afterpage}
\usepackage{capt-of}
\usepackage{graphicx}
\usepackage{enumitem}  

\numberwithin{equation}{section}

\newcommand{\be}{\begin{equation}}
\newcommand{\ee}{\end{equation}}
\newcommand{\non}{\nonumber}
\newcommand{\id}{\mathbb{I}}

\newcommand{\tr}{\mathop{\rm tr}\nolimits}
\newcommand{\str}{\mathop{\rm str}\nolimits}

\begin{document}

\begin{titlepage}
\strut\hfill UMTG--302
\vspace{.5in}
\begin{center}

\LARGE Completing the solution for the $OSp(1|2)$ spin chain\\ 
\vspace{1in}
\large Rafael I. Nepomechie\\
Physics Department, P.O. Box 248046\\
University of Miami, Coral Gables, FL 33124\\[0.8in]
\end{center}

\vspace{.5in}

\begin{abstract}
The periodic $OSp(1|2)$ quantum spin chain has both a graded and a
non-graded version.  Naively, the Bethe ansatz solution for the
non-graded version does not account for the complete spectrum of the
transfer matrix, and we propose a simple mechanism for achieving
completeness.  In contrast, for the graded version, this issue does
not arise.  We also clarify the symmetries of both versions of the
model, and we show how these symmetries are manifested in the
degeneracies and multiplicities of the transfer-matrix spectrum.
While the graded version has $OSp(1|2)$ symmetry, the non-graded 
version has only $SU(2)$ symmetry.
Moreover, we obtain conditions for selecting the physical singular
solutions of the Bethe equations. This analysis solves a lasting 
controversy over signs in the Bethe equations.
\end{abstract}

\end{titlepage}

\setcounter{footnote}{0}

\section{Introduction}\label{sec:intro}

The periodic $OSp(1|2)$ quantum spin chain was first formulated (as a
graded model) and solved in \cite{Kulish:1985bj}.  A non-graded
version of this model was formulated and solved in
\cite{Martins:1994vf, Martins:1994nq}.  These solutions have figured
in various subsequent works, see e.g. \cite{Martins:1995bb,
Ramos:1996np, Sakai:1999nm, Sakai:1999nn, Sakai:2000nd, Saleur:2003zm,
Granet:2018yjg} and references therein.  In this note, we argue that
-- naively -- the solution for the non-graded version does not account
for the complete spectrum of the transfer matrix, and we propose a
simple way to remedy the problem.  In contrast, for the graded
version, this issue does not arise.  We also clarify the symmetries of
both versions of the model, and we show how these symmetries are
manifested in the degeneracies and multiplicities of the
transfer-matrix spectrum. While the graded version has $OSp(1|2)$ symmetry, 
the non-graded version has only $SU(2)$ symmetry.
Moreover, we obtain conditions for
selecting the physical singular solutions of the Bethe equations.  As
a byproduct of our investigation, we clarify the signs in the Bethe
equations, over which there has been some controversy, see e.g.
\cite{Martins:1994nq, Saleur:2003zm}.

We treat the non-graded version of the $OSp(1|2)$ model in Sec.
\ref{sec:nongraded}, and the graded version in Sec.  \ref{sec:graded}.
Our conclusions are in Sec.  \ref{sec:conclusion}. The $OSp(1|2)$ 
symmetry of the graded transfer matrix is proved in the appendix.

\section{The non-graded version}\label{sec:nongraded}

In Sec.  \ref{sec:nongbasics}, we review the construction of the
transfer matrix for the non-graded version of the $OSp(1|2)$ quantum
spin chain, and we note that it has only $SU(2)$ symmetry.  In Sec.
\ref{sec:nongsolution}, we briefly review the solution
\cite{Martins:1994vf, Martins:1994nq}.  The degeneracies and
multiplicities of the transfer-matrix spectrum are explained in Sec.
\ref{sec:nongdegmult}.  We give conditions for selecting the
physical singular solutions of the Bethe equations in Sec.
\ref{sec:nongsingular}.  The difficulty with completeness and its
resolution are discussed in Sec.  \ref{sec:nongcompleteness}.

\subsection{The transfer matrix and its symmetries}\label{sec:nongbasics}

The $OSp(1|2)$ spin chain has a 3-dimensional vector space at each site.
Following \cite{Martins:1994nq}, we consider here the bosonic (non-graded) formulation 
of the model. The R-matrix is given by\footnote{See Eq. (5) in 
\cite{Martins:1994nq} with $\eta=i$.}
\begin{align}
R(u) = 
\left(\begin{array}{ccc|ccc|ccc}
a & 0 & 0 & 0 & 0 & 0 & 0 & 0 & 0 \\
0 & b & 0 & c & 0 & 0 & 0 & 0 & 0 \\
0 & 0 & d & 0 & e & 0 & f & 0 & 0 \\
\hline
0 & c & 0 & b & 0 & 0 & 0 & 0 & 0 \\
0 & 0 & e & 0 & g & 0 & -e & 0 & 0 \\
0 & 0 & 0 & 0 & 0 & b & 0 & c & 0 \\
\hline
0 & 0 & f & 0 & -e & 0 & d & 0 & 0 \\
0 & 0 & 0 & 0 & 0 & c & 0 & b & 0 \\
0 & 0 & 0 & 0 & 0 & 0 & 0 & 0 & a 
\end{array} \right)\,,
\end{align}
where
\begin{align}
a &= i - u \,, \quad b = u \,, \quad c = i \,, \quad d = 
-\frac{u(2u-i)}{2u-3i}\,, \non \\
e &= \frac{2 i u}{2u-3i}  \,, \quad  f = \frac{4 i u +3}{2u-3i} \,, \quad 
g = u + \frac{3}{2u-3i} \,.
\end{align}
It has the regularity property $R(0) = i {\cal P}$, where ${\cal P}$ 
is the (non-graded) permutation matrix on ${\cal C}^{3} \otimes {\cal 
C}^{3}$
\be
{\cal P} = \sum_{a,b=1}^{3} e_{ab} \otimes e_{ba}\,, \quad \left( 
e_{ab} \right)_{ij} = \delta_{a,i} \delta_{b,j} \,.
\label{calP}
\ee
This R-matrix is a solution of the (non-graded) Yang-Baxter equation
\be
R_{12}(u-v)\, R_{13}(u)\, R_{23}(v) = R_{23}(v)\, R_{13}(u)\, 
R_{12}(u-v)\,,
\label{YBE}
\ee
where
\be
R_{12} = R \otimes \id\,, \quad R_{13}={\cal P}_{23}\, R_{12}\, {\cal P}_{23}
\,, \quad R_{23}={\cal P}_{12}\, R_{13}\, {\cal P}_{12} \,.
\label{defs}
\ee
The transfer matrix $t(u)$ for a closed spin chain of length $N$ with periodic 
boundary conditions, which is given as usual by the (non-graded) trace of the 
monodromy matrix
\be
t(u) = \tr_{0} R_{0N}(u)\cdots R_{01}(u) \,,
\label{transfer}
\ee
has the commutativity property
\be
\left[ t(u) \,, t(v) \right] = 0 \,.
\label{commutativity}
\ee
The corresponding Hamiltonian is given by
\be
H = -i \frac{d}{du}\log t(u)\Big\vert_{u=0} \,.
\label{Hamiltonian}
\ee

The transfer matrix has $SU(2)$ symmetry\footnote{The proof is 
similar to the one in Sec. \ref{sec:bosgen} for 
the graded transfer matrix.}
\be
\left[ t(u) \,, S^{z} \right] = \left[ t(u) \,,  S^{\pm} \right] =  0 \,, 
\label{su2transf}
\ee
where
\be
S^{z} = \sum_{n=1}^{N} 
s_{n}^{z} \,, \qquad  
S^{\pm} = \sum_{n=1}^{N} s_{n}^{\pm} \,,
\label{Szng}
\ee
and 
\be
s^{z} = \frac{1}{2}\left(e_{11} - e_{33}\right)\,,  \qquad
s^{+} = e_{13}\,, \qquad
s^{-} = e_{31}\,,
\ee
with $e_{ab}$ defined in (\ref{calP}). 
Indeed, the generators (\ref{Szng}) obey 
the $SU(2)$ algebra 
\be
\left[ S^{z} \,, S^{\pm} \right] = \pm S^{\pm} \,, 
\qquad \left[ S^{+}  \,, S^{-} \right] = 2 S^{z} \,.
\label{algebra}
\ee
 
\subsection{The Bethe ansatz solution}\label{sec:nongsolution}

Let $|\Lambda^{(M)}\rangle$ denote the simultaneous eigenvectors of 
the transfer matrix $t(u)$ (\ref{transfer}) and of $S^{z}$ (\ref{Szng}) 
that are $SU(2)$ highest-weight states\footnote{These vectors can be 
constructed by algebraic Bethe ansatz following \cite{Tarasov:1989}, 
as sketched in appendix B of \cite{Martins:1994nq}. The 
highest-weight property can be checked for $M=1, 2$, and we 
conjecture that it is generally true.\label{footnote:ABA}}
\be
S^{+}|\Lambda^{(M)}\rangle = 0 \,,
\label{hw}
\ee
and let $\Lambda^{(M)}(u)$ and $m$ denote the corresponding
eigenvalues
\begin{align}
t(u)\, |\Lambda^{(M)}\rangle &= \Lambda^{(M)}(u)\, 
|\Lambda^{(M)}\rangle \,, \non \\
S^{z}\, |\Lambda^{(M)}\rangle &= m\, |\Lambda^{(M)}\rangle \,.
\label{simulteig}
\end{align}
It was argued in \cite{Martins:1994nq} that these eigenvalues are given by
\be
m = \frac{1}{2}(N-M) \,, \qquad M = 0, 1, \ldots, N\,,
\label{Szeig}
\ee
and
\begin{align}
\Lambda^{(M)}(u) & = (i-u)^{N} (-1)^{M} 
\frac{Q(u+\frac{i}{2})}{Q(u-\frac{i}{2})} + u^{N}  
\frac{Q(u)\, Q(u-\frac{3i}{2})}{Q(u-i)\, Q(u-\frac{i}{2})} \non\\
& + \left(\frac{u(\frac{i}{2}-u)}{u-\frac{3i}{2}}\right)^{N} (-1)^{M} 
\frac{Q(u-2i)}{Q(u-i)}\,,
\label{TQ}
\end{align}
where
\be
Q(u) = \prod_{k=1}^{M}(u-u_{k}) \,.
\label{Q}
\ee
The corresponding Bethe equations for the Bethe roots $\{u_{1}, 
\ldots, u_{M}\}$ are given by
\be
\left(\frac{u_{j} - \frac{i}{2}}{u_{j} + \frac{i}{2}}\right)^{N} 
= -(-1)^{N-M}\frac{Q(u_{j}-i)\, Q(u_{j}+\frac{i}{2})}{Q(u_{j}+i)\, 
Q(u_{j}-\frac{i}{2})} \,, \qquad j = 1, \ldots, M \,.
\label{BE}
\ee
An unusual feature of the TQ-equation (\ref{TQ}) is its explicit 
dependence (through the factors $(-1)^{M}$) on the number $M$ of 
Bethe roots; and there is a corresponding unusual factor $(-1)^{N-M}$ in the 
Bethe equations (\ref{BE}).
It was implicitly assumed in \cite{Martins:1994nq} that all the Bethe 
roots are finite and pairwise distinct. 

\subsection{Degeneracies and multiplicities}\label{sec:nongdegmult}

The degeneracy (the number of times that a given eigenvalue
$\Lambda^{(M)}(u)$ appears) is given by $N-M+1$, since the eigenstates
form $SU(2)$ irreducible representations with spin $s=m=(N-M)/2$, see 
(\ref{Szeig}), which have dimension $2s+1$.  Since
the 3-dimensional vector space at each site decomposes as spin-1/2 
plus spin-0, the multiplicities in the spectrum follow from the
Clebsch-Gordan decomposition of $({\bf 2} \oplus {\bf
1})^{\otimes N}$. For example,
\begin{align}
N&=2: \qquad \left({\bf 2} \oplus {\bf 
1}\right)^{\otimes 2} 
= 2\cdot {\bf 1} \oplus  2 \cdot{\bf 2} 
\oplus  {\bf 3}  \non\\
N&=3:\qquad \left({\bf 2} \oplus {\bf 1}\right)^{\otimes 3} 
 = 4 \cdot {\bf 1} \oplus  5 \cdot {\bf 2} 
\oplus 3 \cdot {\bf 3} \oplus  {\bf 4} \non\\
N&=4:\qquad \left({\bf 2} \oplus {\bf 1}\right)^{\otimes 4} 
 = 9 \cdot {\bf 1} \oplus  12 \cdot {\bf 2} 
\oplus 9 \cdot {\bf 3} \oplus  4 \cdot {\bf 4} \oplus {\bf 5}
\label{decomp}
\end{align} 
Hence, for $N=2$, there are two eigenvalues with degeneracy 1 ($M=2$),
two eigenvalues with degeneracy 2 ($M=1$), and one eigenvalue with
degeneracy 3 ($M=0$); and similarly for higher $N$. 

\subsection{Physical singular solutions of the Bethe equations}\label{sec:nongsingular}

As in the periodic Heisenberg spin chain \cite{Nepomechie:2013mua, 
Hao:2013jqa, Nepomechie:2014hma}, the Bethe equations 
(\ref{BE}) have many ``singular'' solutions containing $\pm \frac{i}{2}$; only a
subset of these singular solutions, which we call ``physical'',
correspond to actual eigenvalues and eigenstates of the transfer matrix. 
Starting from the Bethe equations for the model with twisted boundary conditions 
(see appendix C in \cite{Martins:1994nq}) and following \cite{Nepomechie:2014hma}, we find that the
condition for the singular solution $\{\frac{i}{2}, -\frac{i}{2}, 
u_{3}, \ldots, u_{M} \}$ to be physical is
\be
\left(-\prod_{j=3}^{M}\frac{u_{j} - \frac{i}{2}}{u_{j} + \frac{i}{2}}\right)^{N} 
= (-1)^{(N-M+1)M} \,.
\label{ngsingular}
\ee
Moreover, $\{u_{3}, \ldots, u_{M} \}$ must also obey
\begin{align}
\left(\frac{u_{j} - \frac{i}{2}}{u_{j} + \frac{i}{2}}\right)^{N-1}
\left(\frac{u_{j} + \frac{3i}{2}}{u_{j} - \frac{3i}{2}}\right)
\left(\frac{u_{j} - i}{u_{j} + i}\right) &= 
(-1)^{N-M+1} \prod_{k=3}^{M} \frac{(u_{j} - u_{k} - i)(u_{j} - u_{k} + 
\frac{i}{2})}{(u_{j} - u_{k} + i)(u_{j} - u_{k} - 
\frac{i}{2})} \,, \non\\
& \qquad\qquad\qquad\qquad j=3, \ldots, M \,.
\label{ngsingularmore}
\end{align}

\subsection{Completing the Bethe ansatz solution}\label{sec:nongcompleteness}

It was argued in Sec.  3 of \cite{Martins:1994nq}, by
considering all finite solutions of the Bethe equations (\ref{BE}) for
the case $N=4$, that the Bethe ansatz solution (\ref{TQ})-(\ref{BE})
accounts for all the eigenvalues of the Hamiltonian
(\ref{Hamiltonian}).

However, we find that this Bethe ansatz solution \emph{cannot} account in this way
for all the eigenvalues of the transfer matrix (\ref{transfer}).  Indeed, for $N=2$,
there is an eigenvalue with $M=1$ that cannot be described by this
solution, namely\footnote{The Bethe equation (\ref{BE}) with $N=2$ and 
$M=1$ has only one finite solution, namely $u_{1}=0$, which describes 
through (\ref{TQ})
a transfer-matrix eigenvalue with $M=1$ that is different from (\ref{missingeigN2}).}
\be
\Lambda^{(1)}(u) = \frac{-4u^{4}+12i u^{3} +29u^{2} - 30 i u 
-9}{(2u-3i)^{2}} \,.
\label{missingeigN2}
\ee
For $N=3$, there are 2 such eigenvalues (with $M=2$); and for $N=4$, 
there are 7 such eigenvalues (1 eigenvalue with $M=1$, and 6 
eigenvalues with $M=3$).

We propose a simple way to account for the missing transfer-matrix eigenvalues:
admit precisely one \emph{infinite} Bethe root. Since the Q-functions
appear in the expression for the eigenvalues (\ref{TQ}) only as ratios, the 
effect in (\ref{TQ}) of one infinite Bethe root ($u_{M} = \infty$) 
is to replace the expression for $Q(u)$ (\ref{Q}) by a product over only the finite roots, i.e.
\be
Q(u) = \prod_{k=1}^{M-1}(u-u_{k}) \,,
\label{Qnew}
\ee
where $\{u_{1}, \ldots, u_{M-1}\}$ are finite. Moreover, the Bethe 
equations for the finite roots are again given by (\ref{BE}) except 
with $Q(u)$ given by (\ref{Qnew}).

We have explicitly verified that all the missing transfer-matrix eigenvalues for 
$N=2,3,4$ can be obtained in this way, with $u_{M} = \infty$. The 
Bethe roots corresponding to all of the transfer-matrix eigenvalues for $N=2,3$ are given in Table 
\ref{table:BRng}. Note that the degeneracies and multiplicities are 
in accordance with the group-theory predictions in Sec. \ref{sec:nongdegmult}.
For $N=4$, we report only the Bethe roots
corresponding to the ``missing'' transfer-matrix eigenvalues, and the 
physical singular solutions, see Table \ref{table:BR}. The physical 
singular solutions in Tables \ref{table:BRng} and \ref{table:BR} 
satisfy (\ref{ngsingular})-(\ref{ngsingularmore}).

\begin{table}[h!]
\centering
\begin{tabular}{cccc}
\hline
$N$ & $M$ & degeneracy & $\{ u_{1}, \ldots u_{M}\}$ \\   
\hline
2 & 0 & 3 &  --  \\
2 & 1 & 2 & $\infty$ \\
2 & 1 & 2 & $0$ \\
2 & 2 & 1 & $\pm \sqrt{2}/4$ \\
2 & 2 & 1 & $\pm i/2$ \\
\hline
3 & 0 & 4 &  --  \\
3 & 1 & 3 &  $\sqrt{3}/2$ \\
3 & 1 & 3 &  $-\sqrt{3}/2$ \\
3 & 1 & 3 &  $0$ \\
3 & 2 & 2 &  $\sqrt{3}/6\,, \infty$\\
3 & 2 & 2 &  $-\sqrt{3}/6\,, \infty$\\
3 & 2 & 2 &  $(3\sqrt{3} \pm i \sqrt{13})/8 $ \\
3 & 2 & 2 &  $(-3\sqrt{3} \pm i \sqrt{13})/8 $ \\
3 & 2 & 2 &  $\pm \sqrt{5}/10$ \\
3 & 3 & 1 &  $0.507442\,, -0.253721 \pm 0.505591 i$ \\
3 & 3 & 1 &  $-0.507442\,, 0.253721 \pm 0.505591 i$ \\
3 & 3 & 1 &  $0\,, \pm 0.550503$\\
3 & 3 & 1 &  $0\,, \pm 0.963355 i$\\
\hline
\end{tabular}
\caption{\small Bethe roots for $N=2, 3$ corresponding to all 
eigenvalues in the non-graded version}\label{table:BRng}
\end{table}

We have also explicitly verified for $N=2,3,4$ that the Bethe states
with $u_{M} = \infty$ and $M= 1, 2$ are indeed eigenvectors of the
transfer matrix.  As already noted in footnote \ref{footnote:ABA}, the
algebraic Bethe ansatz for the $OSp(1|2)$ spin chain is sketched in
appendix B of \cite{Martins:1994nq}.  The Bethe states are constructed
using the creation operators $B_{1}(u)$ and $B_{2}(u)$ defined as in
(\ref{monodromy}).  For Bethe states with an infinite Bethe root, we
use the renormalized creation operator $u^{-N+1} B_{1}(u)$, which is
finite in the limit $u\rightarrow \infty$.

\begin{table}[h!]
\centering
\begin{tabular}{cccc}
\hline
$N$ & $M$ & degeneracy & $\{ u_{1}, \ldots u_{M}\}$ \\   
\hline
4 & 1 & 4 &   $\infty$ \\
4 & 2 & 3 &   $\pm i/2$ \\
4 & 3 & 2 &   $1.06752 \pm 0.421806 i\,, \infty$\\
4 & 3 & 2 &   $-1.06752 \pm 0.421806 i\,, \infty$ \\
4 & 3 & 2 &   $0.0366242\,, 0.431751\,, \infty$\\
4 & 3 & 2 &   $-0.0366242\,, -0.431751\,, \infty$ \\
4 & 3 & 2 &   $\pm 0.422577\,, \infty$ \\
4 & 3 & 2 &   $\pm i/2\,, \infty$\\
4 & 4 & 1 &   $\pm i/2\,, \pm 0.622965$\\
4 & 4 & 1 &   $\pm i/2\,, \pm 1.41883  i$\\
\hline
\end{tabular}
\caption{\small Bethe roots for $N=4$ corresponding to ``missing'' 
eigenvalues, and physical singular solutions, in the non-graded 
version}\label{table:BR}
\end{table}

\section{The graded version}\label{sec:graded}

We turn now to the graded version of the periodic $OSp(1|2)$ quantum
spin chain \cite{Kulish:1985bj}.  We review the construction of the
transfer matrix and its symmetries in Sec.  \ref{sec:gbasics}.  We
review the solution in Sec.  \ref{sec:gsolution}, explain the
degeneracies and multiplicities of the transfer-matrix spectrum in
Sec.  \ref{sec:gdegmult}, and check completeness in Sec.
\ref{sec:gcompleteness}.

\subsection{The transfer matrix and its symmetries}\label{sec:gbasics}

The graded R-matrix is given by\footnote{See Eq. (2.20) in 
\cite{Kulish:1985bj} with $\Delta=\frac{3}{2}$; up to an overall 
factor, our R-matrix is $i R_{\text{Kulish}}(u/i)$.}
\begin{align}
R(u) = 
\left(\begin{array}{ccc|ccc|ccc}
a & 0 & 0 & 0 & 0 & e & 0 & -e & 0 \\
0 & b & 0 & c & 0 & 0 & 0 & 0 & 0 \\
0 & 0 & b & 0 & 0 & 0 & c & 0 & 0 \\
\hline
0 & c & 0 & b & 0 & 0 & 0 & 0 & 0 \\
0 & 0 & 0 & 0 & d & 0 & 0 & 0 & 0 \\
-e & 0 & 0 & 0 & 0 & g & 0 & f & 0 \\
\hline
0 & 0 & c & 0 & 0 & 0 & b & 0 & 0 \\
e & 0 & 0 & 0 & 0 & f & 0 & g & 0 \\
0 & 0 & 0 & 0 & 0 & 0 & 0 & 0 & d 
\end{array} \right)\,,
\label{gRmat}
\end{align}
where
\begin{align}
a &= u+i - \frac{2 i u}{2u-3i} \,, \quad b = u \,, \quad c = i \,, \quad d = 
u-i\,, \non \\
e &= \frac{2 i u}{2u-3i}  \,, \quad  f = -\frac{4 i u +3}{2u-3i} \,, \quad 
g = \frac{u(2u-i)}{2u-3i} \,.
\end{align}
It has the regularity property $R(0) = i {\cal P}$, where ${\cal P}$ 
is now the graded permutation matrix
\be
{\cal P} = \sum_{a,b=1}^{3} (-1)^{p(a)\, p(b)} e_{ab} \otimes e_{ba}\,, 
\ee
where the gradings are given by $p(1)=0\,, p(2) = p(3) =1$. This 
R-matrix is a solution of the graded Yang-Baxter equation, which is 
given by (\ref{YBE}) and (\ref{defs}), but with the graded 
permutation matrix. The transfer matrix $t(u)$ satisfying the 
commutativity property (\ref{commutativity}) is given by 
(\ref{transfer}), but with the factors $R_{0j}(u)$ constructed as in 
(\ref{defs}) using the graded permutation matrix, and
with the graded trace (supertrace) $\str X = 
\sum_{a=1}^{3} (-1)^{p(a)} X_{aa}$.

The transfer matrix now has $OSp(1|2)$ symmetry, in contrast with the 
non-graded version that has only $SU(2)$ symmetry. Indeed, we show 
in the appendix that
\be
\left[ t(u) \,, S^{z} \right] = \left[ t(u) \,,  S^{\pm} \right] =  
\left[ t(u) \,,  J^{\pm} \right] = 0 \,, 
\label{osp12transf}
\ee
where the generators are given by
\be
S^{z} = \sum_{n=1}^{N} 
s_{n}^{z} \,, \qquad  
S^{\pm} = \sum_{n=1}^{N} s_{n}^{\pm} \,, \qquad
J^{\pm} = \sum_{n=1}^{N} j_{n}^{\pm} P_{n+1} \cdots P_{N}\,, 
\label{gens}
\ee
with
\be
s^{z} = \frac{1}{2}\left(e_{33} - e_{22}\right)\,, \qquad
s^{+} = e_{32}\,, \qquad s^{-} = e_{23} \,,
\label{onesitespin}
\ee
and
\be
j^{+} = -e_{12} - e_{31} \,, \qquad  j^{-} = -e_{13} + e_{21} \,, \qquad
P = e_{11} - e_{22} -e_{33}\,.
\label{onesitefermi}
\ee
These generators satisfy the $OSp(1|2)$ algebra
\begin{align}
\left[ S^{z} \,, S^{\pm} \right] &= \pm S^{\pm} \,, 
\qquad 
\left[ S^{z} \,, J^{\pm} \right] = \pm \frac{1}{2}J^{\pm} \,, 
\qquad \left[ S^{+}  \,, S^{-} \right] = 2 S^{z} \,, \non \\
\left\{ J^{+} \,, J^{-} \right\} &=  2 S^{z} \,, 
\qquad 
\left\{ J^{\pm} \,, J^{\pm} \right\} = \pm 2 S^{\pm} \,, 
\qquad \left[ S^{\pm}  \,, J^{\mp} \right] = - J^{\pm} \,,
\label{ospalgebra}
\end{align}
and all other (anti-)commutators vanish.
Note that the fermionic generators $J^{\pm}$ have 
a non-trivial coproduct (\ref{gens}) involving the grading involution 
$P$, see e.g. \cite{Tsujimoto2011, Crampe:2019}.

\subsection{The Bethe ansatz solution}\label{sec:gsolution}

Let us denote by $|\Lambda^{(M)}\rangle$ the simultaneous
eigenvectors of $t(u)$ and $S^{z}$ that are $OSp(1|2)$ highest-weight
states\footnote{As in the non-graded version, the highest-weight 
property can be checked for small values of $M$, and is conjectured to be 
generally true.}
\be
S^{+}|\Lambda^{(M)}\rangle = 0 \,, \qquad J^{+}|\Lambda^{(M)}\rangle = 0 \,,
\label{ghw}
\ee
with corresponding eigenvalues $\Lambda^{(M)}(u)$ and $m$ 
\begin{align}
t(u)\, |\Lambda^{(M)}\rangle &= \Lambda^{(M)}(u)\, 
|\Lambda^{(M)}\rangle \,, \non \\
S^{z}\, |\Lambda^{(M)}\rangle &= m\, |\Lambda^{(M)}\rangle \,.
\label{gsimulteig}
\end{align}
These eigenvalues are given by \cite{Kulish:1985bj}
\be
m = \frac{1}{2}(N-M) \,, \qquad M = 0, 1, \ldots, N\,,
\label{gSzeig}
\ee
and
\begin{align}
\Lambda^{(M)}(u) = -(u-i)^{N} 
\frac{Q(u+\frac{i}{2})}{Q(u-\frac{i}{2})} + u^{N}  
\frac{Q(u)\, Q(u-\frac{3i}{2})}{Q(u-i)\, Q(u-\frac{i}{2})} 
- \left(\frac{u(u-\frac{i}{2})}{u-\frac{3i}{2}}\right)^{N} 
\frac{Q(u-2i)}{Q(u-i)}\,,
\label{gTQ}
\end{align}
where $Q(u)$ is again given by (\ref{Q}). The corresponding Bethe 
equations are given by
\be
\left(\frac{u_{j} - \frac{i}{2}}{u_{j} + \frac{i}{2}}\right)^{N} 
= \frac{Q(u_{j}-i)\, Q(u_{j}+\frac{i}{2})}{Q(u_{j}+i)\, 
Q(u_{j}-\frac{i}{2})} \,, \qquad j = 1, \ldots, M \,.
\label{gBE}
\ee
In contrast with the non-graded version, the TQ-equation and Bethe 
equations 
do \emph{not} depend explicitly on the number $M$ of Bethe roots.

We remark that the reference state (i.e., the state $|\Lambda^{(M)}\rangle$ 
with $M=0$) in the graded version is 
given by $\left(\begin{smallmatrix}
0\\ 0 \\ 1
\end{smallmatrix}\right)^{\otimes N}$, while in the non-graded version it is given by 
$\left( \begin{smallmatrix}
1 \\ 0 \\ 0 \end{smallmatrix} \right)^{\otimes N}$.

The condition for the singular solution $\{\frac{i}{2}, -\frac{i}{2}, 
u_{3}, \ldots, u_{M} \}$ to be physical is now
\be
\left(-\prod_{j=3}^{M}\frac{u_{j} - \frac{i}{2}}{u_{j} + \frac{i}{2}}\right)^{N} 
= 1 \,,
\label{gsingular}
\ee
together with
\be
\left(\frac{u_{j} - \frac{i}{2}}{u_{j} + \frac{i}{2}}\right)^{N-1}
\left(\frac{u_{j} + \frac{3i}{2}}{u_{j} - \frac{3i}{2}}\right)
\left(\frac{u_{j} - i}{u_{j} + i}\right) = 
\prod_{k=3}^{M} \frac{(u_{j} - u_{k} - i)(u_{j} - u_{k} + 
\frac{i}{2})}{(u_{j} - u_{k} + i)(u_{j} - u_{k} - 
\frac{i}{2})}\,, \quad j=3, \ldots, M \,,
\label{gsingularmore}
\ee
cf. (\ref{ngsingular})-(\ref{ngsingularmore}).

\subsection{Degeneracies and multiplicities}\label{sec:gdegmult}

The degeneracy (the number of times that a given eigenvalue
$\Lambda^{(M)}(u)$ appears) is given by $2N-2M+1$, since the eigenstates
form $OSp(1|2)$ irreducible representations with spin $s=m=(N-M)/2$, see 
(\ref{gSzeig}), which have dimension $4s+1$. Indeed, these irreps, 
which we denote by $[s]$, consist of a \emph{pair} of $SU(2)$ irreps that 
have highest weights $|\Lambda^{(M)}\rangle$ (with spin $s$ and dimension $2s+1$) and 
$J^{-}|\Lambda^{(M)}\rangle$ (with spin $s-\frac{1}{2}$ and dimension 
$2s$).

Since the 3-dimensional vector space at each site forms an irrep $[\frac{1}{2}]$, the multiplicities 
in the spectrum follow from the $OSp(1|2)$ decomposition of 
$[\frac{1}{2}]^{\otimes N}$, which can be easily computed from the 
fact (see e.g. \cite{Rasmussen:2002rx} and references therein)
\be
[s_{1}] \otimes [s_{2}] = [|s_{1}-s_{2}|] \oplus 
[|s_{1}-s_{2}|+\frac{1}{2}] \oplus \cdots \oplus 
[s_{1}+s_{2}-\frac{1}{2}] \oplus [s_{1}+s_{2}] \,.
\ee 
For example, in terms of the dimensions of the irreps,
\begin{align}
N&=2: \qquad {\bf 3}^{\otimes 2} 
= {\bf 1} \oplus {\bf 3} 
\oplus  {\bf 5}  \non\\
N&=3:\qquad {\bf 3}^{\otimes 3} 
 = {\bf 1} \oplus  3 \cdot {\bf 3} 
\oplus 2 \cdot {\bf 5} \oplus  {\bf 7} \non\\
N&=4:\qquad {\bf 3}^{\otimes 4} 
 = 3 \cdot {\bf 1} \oplus  6 \cdot {\bf 3} 
\oplus 6 \cdot {\bf 5} \oplus  3 \cdot {\bf 7} \oplus {\bf 9}
\label{gdecomp}
\end{align} 
Hence, for $N=2$, there is one eigenvalue with degeneracy 1 ($M=2$),
one eigenvalue with degeneracy 3 ($M=1$), and one eigenvalue with
degeneracy 5 ($M=0$); and similarly for higher $N$. Evidently, the 
patterns (\ref{gdecomp}) differ significantly from those in 
the non-graded version (\ref{decomp}).

\subsection{Checking completeness}\label{sec:gcompleteness}

We have explicitly verified for $N=2, 3, 4$ that the Bethe ansatz
solution (\ref{gSzeig})-(\ref{gBE}) accounts for all the eigenvalues
of the transfer matrix, where all the Bethe roots are finite and pairwise
distinct. In contrast with the non-graded version, an infinite Bethe root is \emph{not}
necessary, which is consistent with the fact that the TQ-equation and 
Bethe equations do not depend explicitly on $M$.

The Bethe roots for $N=2, 3, 4$ are given in Table \ref{table:BRg}.
Note that the degeneracies and multiplicities are 
in accordance with the group-theory predictions in Sec. \ref{sec:gdegmult}.
The physical singular solutions in Table \ref{table:BRg} 
satisfy (\ref{gsingular})-(\ref{gsingularmore}).

\begin{table}[h!]
\centering
\begin{tabular}{cccc}
\hline
$N$ & $M$ & degeneracy & $\{ u_{1}, \ldots u_{M}\}$ \\   
\hline
2 & 0 & 5 &  --  \\
2 & 1 & 3 & $0$ \\
2 & 2 & 1 & $\pm i/2$ \\
\hline
3 & 0 & 7 &  --  \\
3 & 1 & 5 &  $\sqrt{3}/6$ \\
3 & 1 & 5 &  $-\sqrt{3}/6$ \\
3 & 2 & 3 &  $(3\sqrt{3} \pm i \sqrt{13})/8 $\\
3 & 2 & 3 &  $(-3\sqrt{3} \pm i \sqrt{13})/8$\\
3 & 2 & 3 &  $\pm \sqrt{5}/10$ \\
3 & 3 & 1 &  $0\,, \pm i/2$ \\
\hline
4 & 0 & 9 &  --  \\
4 & 1 & 7 & $1/2$ \\
4 & 1 & 7 & $-1/2$ \\
4 & 1 & 7 & $0$ \\
4 & 2 & 5 & $1.06752 \pm 0.421806 i$ \\
4 & 2 & 5 & $-1.06752 \pm 0.421806 i$ \\
4 & 2 & 5 & $0.0366242\,, 0.431751$ \\
4 & 2 & 5 & $-0.0366242\,, -0.431751$ \\
4 & 2 & 5 & $\pm i/2$ \\
4 & 2 & 5 & $\pm \sqrt{5/28}$ \\
4 & 3 & 3 & $0.106997\,, 0.696501 \pm 0.464584 i$ \\
4 & 3 & 3 & $-0.106997\,, -0.696501 \pm 0.464584 i$ \\
4 & 3 & 3 & $0.367029\,, -0.795887 \pm 0.448098 i$ \\
4 & 3 & 3 & $-0.367029\,, 0.795887 \pm 0.448098 i$ \\
4 & 3 & 3 & $0\,, \pm 1.01641 i$ \\
4 & 3 & 3 & $0\,, \pm 0.364822 $\\
4 & 4 & 1 & $\pm 1.67174 i\,, \pm i/2$ \\
4 & 4 & 1 & $\pm 0.211488\,, \pm i/2$ \\
4 & 4 & 1 & $0.589529 \pm 0.471746 i\,, -0.589529 \pm 0.471746 i$ \\
\hline
\end{tabular}
\caption{\small Bethe roots for $N=2, 3, 4$ corresponding to all 
eigenvalues in the graded version}\label{table:BRg}
\end{table}

\section{Conclusions}\label{sec:conclusion}

We have argued that, by allowing for the possibility of an infinite
Bethe root (and all other Bethe roots finite and pairwise distinct),
the Bethe ansatz solution for the non-graded version of the periodic
$OSp(1|2)$ spin chain (\ref{TQ})-(\ref{BE}) \emph{can} account for all
the distinct eigenvalues of the transfer matrix (\ref{transfer}).

We emphasize the difference with respect to, say, the Bethe ansatz 
solution for the periodic $SU(2)$-invariant Heisenberg (XXX) spin chain, which (as far 
as has been checked \cite{Hao:2013jqa}) accounts for 
all the transfer-matrix eigenvalues by means of \emph{finite} Bethe 
roots.\footnote{As is well known, in the algebraic Bethe ansatz approach 
for the Heisenberg spin chain, the Bethe states are given by 
$B(u_{1}) \cdots B(u_{M})|0\rangle$, where $B(u)$ is a certain 
creation operator and $|0\rangle$ is a reference state. If all the Bethe roots 
$\{ u_{1}, \ldots, u_{M}\}$ are finite, then the Bethe state is an
$SU(2)$ highest-weight state; and the lower-weight states (which have 
the same transfer-matrix eigenvalue as the Bethe state)
can be obtained by repeatedly acting with the spin-lowering operator $S^{-}$ on the 
Bethe state. Since $B(\infty) \sim S^{-}$, some authors prefer to 
describe such lower-weight states of the Heisenberg spin chain
in terms of (multiple) infinite Bethe 
roots. In contrast, for the non-graded version of the $OSp(1|2)$ spin chain, 
we find (as noted at the end of Sec. \ref{sec:nongcompleteness})
that one infinite Bethe root is necessary to construct certain \emph{Bethe states}.}

The reason that, in the non-graded version, an infinite Bethe root can
help give a ``missing'' transfer-matrix eigenvalue (i.e., an eigenvalue
that cannot be obtained with exclusively finite Bethe roots) is that
the TQ-equation depends explicitly on the number $M$ of Bethe roots.
Indeed, if a (homogeneous) TQ-equation does not depend explicitly on $M$, then
taking one Bethe root to infinity simply gives the \emph{same}
TQ-equation with one less Bethe root, so nothing new can be obtained.
We expect that a similar mechanism may be necessary for achieving
completeness of the transfer-matrix spectrum in other integrable
models whose TQ-equation depends explicitly on the number of Bethe
roots, e.g. \cite{Martins:1995bb}.

By verifying the completeness of the Bethe ansatz solutions for both
the non-graded and graded versions of the model for small values of
$N$, we have seen that both Bethe equations (\ref{BE}) and (\ref{gBE})
appear to correctly describe the corresponding spectra.  However, for
the non-graded version, $SU(2)$ symmetry alone completely accounts for
the degeneracies and multiplicities of the transfer-matrix spectrum.
Only the graded transfer matrix has $OSp(1|2)$ symmetry, which is
reflected in the degeneracies and multiplicities of its spectrum.
This result is in agreement with the observation in
\cite{Saleur:2003zm}, and in disagreement with \cite{Martins:1994nq},
that the Bethe equations for the model with $OSp(1|2)$ symmetry are
(\ref{gBE}), and not (\ref{BE}).

The thermodynamic limit and finite-size corrections
of the non-graded model, which does not have $OSp(1|2)$ symmetry, have
been investigated in \cite{Martins:1994nq, Sakai:1999nm, 
Sakai:1999nn, Sakai:2000nd}. It may be worth checking whether the existence of an
infinite Bethe root noted here changes those results.
Moreover, it is possible that the
thermodynamic limit and finite-size corrections of the graded model,
which does have $OSp(1|2)$ symmetry, could be different.  However, to 
our knowledge, this remains to be investigated. 

\section*{Acknowledgments}
I thank Etienne Granet for bringing the $OSp(1|2)$ model to my 
attention, and for reading the draft.

\appendix

\section{Proof of $OSp(1|2)$ symmetry}\label{sec:proof}

We show here that the graded transfer matrix has $OSp(1|2)$ symmetry 
(\ref{osp12transf}). We first consider the bosonic generators $S^{z}\,, S^{\pm}$ in 
Sec. \ref{sec:bosgen}, and then consider the fermionic generators 
$J^{\pm}$ in Sec. \ref{sec:fermgen}. For later reference, we
introduce the monodromy matrix
\be
T_{0}(u) = R_{0N}(u)\cdots R_{01}(u) = \begin{pmatrix}
A_{1}(u) & B_{1}(u)  & B_{2}(u) \\
C_{1}(u) & A_{2}(u)  & B_{3}(u) \\
C_{2}(u) & C_{3}(u)  & A_{3}(u) 
\end{pmatrix} \,,
\label{monodromy}
\ee
in terms of which the transfer matrix is given by 
\be
t(u) = \str_{0} T_{0}(u) = A_{1}(u) - A_{2}(u)  - A_{3}(u) \,.
\label{transf}
\ee

\subsection{Bosonic generators}\label{sec:bosgen}

The proof for the bosonic generators is similar to the classic proof 
of $SU(2)$ symmetry for the Heisenberg spin chain \cite{Faddeev:1981ft}.
The $SU(2)$ symmetry of the R-matrix (\ref{gRmat}) means that 
\be
\left[\vec s_{1}\,, R_{12}(u) \right] = -\left[\vec s_{2}\,, R_{12}(u) 
\right] \,,
\label{Rmatsu2}
\ee
where $\vec s = (s^{x}, s^{y}, s^{z})$ are the 1-site spin operators, 
see (\ref{onesitespin}). 
Using the fact $\vec S = \sum_{n=1}^{N} \vec s_{n}$, we obtain
{\allowdisplaybreaks 
\begin{align}
\left[\vec S\,, T_{0}(u) \right] &= \sum_{n=1}^{N} \left[\vec 
s_{n}\,, T_{0}(u) \right] 
= \sum_{n=1}^{N} \left[\vec s_{n}\,, R_{0N}(u)\cdots R_{01}(u) \right] \non \\
&= \sum_{n=1}^{N} \sum_{k=1}^{N} R_{0N}(u)\cdots \left[\vec s_{n}\,,  
R_{0k}(u) \right] \cdots R_{01}(u)  \non \\
&= \sum_{n=1}^{N}  R_{0N}(u)\cdots \left[\vec s_{n}\,,  
R_{0n}(u) \right] \cdots R_{01}(u)  \non \\
&= - \sum_{n=1}^{N}  R_{0N}(u)\cdots \left[\vec s_{0}\,,  
R_{0n}(u) \right] \cdots R_{01}(u) \,,
\end{align}}
where we have used (\ref{Rmatsu2}) to pass to the final line.  
We therefore arrive at the identity
\be
\left[\vec S\,, T_{0}(u) \right] = - \left[\vec s_{0}\,, T_{0}(u) 
\right] \,.
\label{identity}
\ee
The RHS of (\ref{identity}) can be readily evaluated using the 
expressions (\ref{onesitespin}) for the 1-site spin operators and 
the expression (\ref{monodromy}) for the monodromy matrix. In 
particular, we obtain
\begin{align}
\left[S^{z} \,, A_{i}(u) \right] &= 0 \,, \qquad i = 1, 2, 3\,,  \non \\
\left[S^{+} \,, A_{1}(u) \right] &= 0 \,, \qquad  
\left[S^{+} \,, A_{2}(u) \right] = B_{3}(u) \,, \qquad  
\left[S^{+} \,, A_{3}(u) \right] = -B_{3}(u) \,, \non \\
\left[S^{-} \,, A_{1}(u) \right] &= 0 \,, \qquad
\left[S^{-} \,, A_{2}(u) \right] = -C_{3}(u) \,, \qquad  
\left[S^{-} \,, A_{3}(u) \right] = C_{3}(u) \,.
\end{align}
In view of the expression (\ref{transf}) for the transfer matrix, we 
conclude that
\be
\left[S^{z} \,, t(u) \right] = 0\,, \qquad \left[S^{\pm} \,, t(u) 
\right]=0 \,.
\ee

\subsection{Fermionic generators}\label{sec:fermgen}

The proof for the fermionic generators $J^{\pm}$ is similar, except
that the coproduct is no longer trivial (\ref{gens}). We start from
the symmetry of the R-matrix (\ref{gRmat})
\be
\left[j^{\pm}_{1} P_{2}\,, R_{12}(u) \right] = -\left[j^{\pm}_{2}\,, R_{12}(u) 
\right] \,,
\label{Rmatosp}
\ee
where the 1-site operators are defined in (\ref{onesitefermi}). 
Proceeding as before, we obtain a result analogous to (\ref{identity})
\be
\left[J^{\pm}\,, T_{0}(u) \right] = - \left[j^{\pm}_{0}\Pi \,, T_{0}(u) 
\right] \,,
\label{ospidentity}
\ee
where we have introduced the quantum-space operator
\be
\Pi = P_{1} \cdots P_{N} \,.
\ee 
This operator commutes with the monodromy matrix's bosonic elements
\be
\left[\Pi \,, A_{i}(u) \right] = 0  \,, \quad i = 1, 2, 3\,, \qquad 
\left[\Pi \,, B_{3}(u) \right] = 0  \,, \qquad \left[\Pi \,, C_{3}(u) 
\right] = 0 \,, 
\ee
and anticommutes with its fermionic elements
\be
\left\{\Pi \,, B_{i}(u) \right\} = 0 \,, \qquad 
\left\{\Pi \,, C_{i}(u) \right\} = 0 \,, \qquad i = 1, 2 \,.
\ee
It is now straightforward to evaluate the RHS of (\ref{ospidentity}), 
and we obtain
\begin{align}
\left[J^{+} \,, A_{1}(u) \right] &= -\left(B_{2}(u)+ C_{1}(u)\right) 
\Pi \,, \quad  
\left[J^{+} \,, A_{2}(u) \right] = -C_{1}(u) \Pi \,, \quad  
\left[J^{+} \,, A_{3}(u) \right] = -B_{2}(u) \Pi \,, \non \\
\left[J^{-} \,, A_{1}(u) \right] &= \left(B_{1}(u) - C_{2}(u)\right) 
\Pi \,, \quad  
\left[J^{-} \,, A_{2}(u) \right] =  B_{1}(u) \Pi \,, \quad  
\left[J^{-} \,, A_{3}(u) \right] = -C_{2}(u) \Pi \,.
\end{align}
In view of the expression (\ref{transf}) for the transfer matrix, we 
conclude that
\be
\left[J^{\pm} \,, t(u) \right] = 0 \,.
\ee

% \bibliographystyle{utphys}
% \bibliography{refs}

\providecommand{\href}[2]{#2}\begingroup\raggedright\endgroup

\end{document}